\documentclass[12pt,onecolumn]{IEEEtran}
\usepackage{graphicx,psfrag,epsfig,epsf,latexsym,hhline,amsmath,amssymb,multirow}
\usepackage[usenames,dvipsnames]{pstricks}
\usepackage{pst-plot}
\usepackage{pstricks-add}
\usepackage{graphicx}
\usepackage{amsthm}
\usepackage{algorithm}
\usepackage{algpseudocode}

\usepackage{blindtext}
\usepackage{etoolbox}
\usepackage{color}

\newcommand{\mc}[1]{\mathcal{#1}}

\newcommand{\mbb}[1]{\mathbb{#1}}

\newcommand{\defeq}{\triangleq}
\newcommand{\Pp}{\mathbb{P}}

\newcommand{\iid}{i.\@i.\@d.\ }

\theoremstyle{definition}\newtheorem{lemma}{Lemma}
\theoremstyle{definition}
\theoremstyle{definition}\newtheorem{theorem}[lemma]{Theorem}
\theoremstyle{definition}
\newtheorem{define}[lemma]{Definition}
\newtheorem{example}[lemma]{Example}

\newtheorem{remark}[lemma]{Remark}

\newcommand\shortintertext[1]{%
\ifvmode\else\\\@empty\fi
\noalign{%
\penalty0%
\vbox{\mathstrut}%
\penalty10000%
\vskip-\baselineskip
\penalty10000%
\vbox to 0pt{%
\normalbaselines
\ifdim\linewidth=\columnwidth
\else
\parshape\@ne
\@totalleftmargin\linewidth
\fi
\vss
\noindent#1\par}%
\penalty10000%
\vskip-\baselineskip}%
\penalty10000}

\begin{document}
\title{Asynchronous Physical-Layer Network Coding with Quasi-Cyclic Codes}
\author{Ping-Chung Wang, Yu-Chih Huang, and Krishna R. Narayanan,\\
Department of Electrical and Computer Engineering \\
Texas A\&M University\\
{\tt\small {\{michael422603@tamu.edu, jerry.yc.huang@gmail.com, krn@ece.tamu.edu\}} }
\thanks{Part of the results in this paper has been submitted to the 2014 IEEE Global Communications Conference \cite{michael14}.}}

\date{}
\maketitle

\begin{abstract}
Communication in the presence of bounded timing asynchronism which is known to the receiver but cannot be easily compensated is studied. Examples of such situations include point-to-point communication over inter-symbol interference (ISI) channels and asynchronous wireless networks. In these scenarios, although the receiver may know all the delays, it is often not be an easy task for the receiver to compensate the delays as the signals are mixed together. A novel framework called interleave/deinterleave transform (IDT) is proposed to deal with this problem. It is shown that the IDT allows one to \textit{design} the delays so that quasi-cyclic (QC) codes with a proper shifting constraint can be used accordingly. When used in conjunction with QC codes, IDT provides significantly better performance than existing schemes relying solely on cyclic codes. Two instances of asynchronous physical-layer network coding, namely the integer-forcing equalization for ISI channels and asynchronous compute-and-forward, are then studied. For integer-forcing equalization, the proposed scheme provides improved performance over using cyclic codes. For asynchronous compute-and-forward, the proposed scheme shows that there is no loss in the achievable information due to delays which are integer multiples of the symbol duration. Further, the proposed approach shows that delays introduced by the channel can sometimes be exploited to obtain higher information rates than those obtainable in the synchronous case. The proposed IDT can be thought of as a generalization of the interleaving/deinterleaving idea proposed by Wang et al. which allows the use of QC codes thereby substantially increasing the design space.
\end{abstract}



\section{Introduction}

Physical-layer network coding (or compute-and-forward) \cite{zhang06} \cite{nazer_tutorial} \cite{nazer2011CF} has been shown to be a way to effectively harness interference in wireless networks and to provide significantly higher throughput than conventional strategies for many wireless networking problems. However, most of the results in the literature consider the case when the time delays from the multiple transmitters to a receiver are all identical (we refer to this as the synchronous case). One of the important open problems in this area is to determine whether the information rates achieved with compute-and-forward in the synchronous case can be obtained when the time delays from the multiple transmitters are different also (we refer to this as the asynchronous case). So far, this question has not been conclusively answered and our understanding of asynchronous physical-layer network coding is not as thorough as that of synchronous one. Recently, there have been some efforts in the literature trying to address such problems for some specific models such as ISI \cite{ordentlich12} and asynchronous physical-layer network coding (and compute-and-forward as well) \cite{FuShenli09} \cite{Lu12} \cite{wu13} \cite{najafi13}. In both cases, cyclic codes have been suggested for combating the time delays for these two seemingly different problems \cite{ordentlich12} \cite{wu13}. While cyclic codes are quite useful for these problems, there has been no proof that rates achievable in the synchronous case are achievable in the asynchronous case also. One important reason for this is the fact that there is no proof showing the existence of ensembles of cyclic codes that can achieve capacity.

In this paper, we show that there is no fundamental loss in the achievable information rates due to asynchronism, when the time delays introduced by the channel are integer multiples of a symbol duration. Interestingly, we also show that in some scenarios, time delays introduced by the channel can be exploited to achieve higher information rates than those achievable in the synchronous case. These results are based on two insights and novel ideas proposed in this paper. The first is the insight that cyclic codes are \textit{not necessary} to deal with asynchronism and that quasi-cyclic (QC) codes suffice. The second main idea is to use an interleve/deinterleave transform (IDT) which equips QC codes with the capability to combat time delays. With a slight rate reduction, this transform will convert any linear shift of an integer multiple of the symbol duration introduced by the channel into a circular shift of another integer value depending on the parameter we choose. Therefore, one can then utilize the IDT for \textit{designing} the equivalent time delays seen at the transform output and implement a QC code accordingly. We then show the existence of an ensemble of QC codes that can achieve capacity for channels whose capacity can be achieved by linear codes and leverage this result to prove the aforementioned information-theoretic result for asynchronous physical layer network coding.


To give concrete examples, we implement the proposed IDT together with QC codes for two instances of asynchronous physical-layer network coding, namely integer-forcing equalization for ISI channels and asynchronous compute-and-forward. For the integer-forcing equalization proposed in \cite{ordentlich12}, we show that our IDT-QC codes achieve the upper bound on information rates presented in \cite{ordentlich12} which may not be achievable by the cyclic coding scheme proposed therein. For asynchronous compute-and-forward, when the delays are integer multiples of the symbol duration, we first show that the rates achievable in the synchronous case can also be achieved in the asynchronous case. In addition to this, we also show that the proposed IDT-QC codes are capable of exploiting another dimension, namely the delay dimension which leads to rates exceeding those achieved in synchronous compute-and-forward \cite{nazer2011CF}. Finally, we consider the case of non-integer valued delays and when rectangular pulses are used, we show that the proposed schemes achieves higher rates than the scheme in \cite{najafi13}.  It is worth noting that the proposed IDT-QC codes are not limited to these two specific examples and can potentially be implemented for many networks with delays which cannot be easily compensated.

In addition to being of theoretical importance, the use of quasi-cyclic (QC) codes is of substantial practical importance as well. QC codes, QC low-density parity check (LDPC) codes in particular, are quite popular in modern coding theory due to their following desirable properties.  They can be encoded using linear feedback shift registers \cite{qcenc1} and a message passing decoder can be implemented efficiently in hardware in a partially parallel architecture \cite{qcdec1}. Further, the QC property makes it efficient to route wires when implementing the message passing decoder \cite{qcdec2}. Moreover, the family of QC codes is much larger than and subsumes as a special case the family of cyclic codes. Due to these properties, QC LDPC codes have been included in many real world applications such as IEEE 802.11n \cite{80211n}, IEEE 802.16e \cite{80216e}, DVB-S2 \cite{DVBS2}, etc. In this paper, we show that in addition to these desirable properties, when used with the IDT transform, QC codes can be a perfect candidate for combating time delays.

The proposed IDT framework can be regarded as a generalization of the scheme in \cite{FuShenli09} where a pair of interleaver/deinterleaver has been implemented together with convolutional codes. In the very last stage of the preparation of this paper, we became aware of a very recently posted independent work \cite{YangLiew13} where an idea similar to \cite{FuShenli09} has been used together with tail-biting convolutional codes for asynchronous physical-layer network-coding for the two-way relay channel. Our paper differs from \cite{FuShenli09} and \cite{YangLiew13} in the following two important ways. Firstly, in contrast to \cite{FuShenli09} and \cite{YangLiew13} which consider only convolutional codes and cyclic codes, our generalization permits the use of any QC linear/lattice code, thereby expanding the design space for the codes that can be used with asynchronism. Secondly, the use of QC codes allows us to derive capacity results for channels with asynchronism.

\subsection{Organization}
The paper is organized as follows. In Section~\ref{sec:prelim}, we provide definitions of cyclic codes and QC codes and also review a well-known construction of QC codes based on protographs. We also review the modulation scheme typically used in the compute-and-forward literature. The modulation scheme has been shown to preserve the structures induced by the channel and hence are crucial for compute-and-forward. This review is of practical importance as our proposed scheme heavily relies on QC codes and the modulation scheme and the family of QC LDPC codes is one of the most popular classes of QC codes in practice. In Section~\ref{sec:proposed_framework}, we elucidate the proposed IDT-QC codes and show some properties of the proposed codes which include the capacity-achieving property. Section~\ref{sec:IF_ISI} and Section~\ref{sec:asyn_CF} provide two interesting applications of the proposed IDT-QC codes, namely point-to-point communication over ISI channels and asynchronous compute-and-forward. In Section~\ref{sec:practical}, we introduce a new joint detection and decoding scheme for our proposed IDT-QC codes. This section lifts the information-theoretic framework proposed in Section~\ref{sec:proposed_framework}-\ref{sec:asyn_CF} towards practical implementation by explicitly introducing a practically implementable decoding scheme. Finally, Section~\ref{sec:conclude} concludes the paper.


\subsection{Notation}
Throughout the paper, $\mbb{R}$, $\mbb{C}$, and $\mbb{Z}$ represent the set of real numbers, complex numbers, and integers, respectively. $\Pp(\mathcal{E})$ denotes the probability of the event $\mathcal{E}$. Vectors and matrices are written in lowercase boldface and uppercase boldface, respectively. We use $*$ to denote linear convolution. For a vector $\mathbf{x}$, we use $\mathbf{x}^{(t)}$ to denote the right circularly shifted version of $\mathbf{x}$ by $t$. e.g., for $\mathbf{x}=[1,2,3,4]$, $\mathbf{x}^{(1)} = [4,1,2,3]$. Moreover, $\oplus$ and $\odot$ are addition and multiplication, respectively, over a finite field whose size is understood from the context.

\section{Preliminaries}\label{sec:prelim}

We first give definitions of cyclic codes and QC codes and then discuss a well-known construction of QC LDPC codes.

\begin{define}[Cyclic codes]
    A linear code $\mc{C}$ is a cyclic code of length $N$ if any circular shift of a codeword is a codeword in $\mc{C}$, i.e., for every $\mathbf{c} \in \mc{C}$, $\mathbf{c}^{(i)} \in \mc{C}$, for all $i=0,\ldots,N-1$.
\end{define}

\begin{define}[Quasi-cyclic codes - Representation I]
    A linear code $\mc{C}$ is a QC code with shifting constraint $b$ if any circular shift of a codeword by a multiple of $b$ is a codeword in $\mc{C}$, i.e., for every $\mathbf{c} \in \mc{C}$, $\mathbf{c}^{(bi)} \in \mc{C}$, for all $i=0,\ldots,\left \lfloor{\frac{N}{b}}\right \rfloor -1$.
    \label{def:repre1}
\end{define}

\begin{define}[Quasi-cyclic codes - Representation II]
    A linear code $\mc{C}$ is a QC code with shifting constraint $b$ if every codeword $\mathbf{c}\in\mc{C}$ consists of $b$ sub-blocks and for each codeword, circularly shifting every sub-block by the same amount results in a codeword.
\end{define}
Note that the above two representations of QC codes are equivalent and such codes are referred to as $b$-QC codes. One can be converted to the other via an interleaver. Throughout the paper, unless mentioned otherwise, the first representation of QC codes is adopted (Definition \ref{def:repre1}). On the other hand, many constructions in the literature (e.g. \cite{Lan07,CCSD}) adopt the second representation.

LDPC codes have been very popular in modern coding theory and in practice due to its ability to achieve near-capacity performance with low decoding complexity and outstanding performance in the finite-length regime. The family of QC LDPC codes is a special class of LDPC codes possessing the QC property that have efficient encoding and decoding algorithms. In what follows, we briefly review the construction of QC LDPC codes. Most of the works in the literature consider using the protograph-based construction of \cite{Thorpe} to generate QC LDPC codes, see for example \cite{Lan07,CCSD} and the reference therein.

To construct a protograph of a length $N$ $b$-QC LDPC code, one begins with a $c\times b$ protomatrix and then replaces each entry in the protomatrix by an $L\times L$ matrix where $L\defeq N/b$. The replacement follows the rule that if the entry is 1, it is replaced by a random $L\times L$ permutation matrix while if the entry is 0, it is replaced by an all-zero matrix. This would result in a $cL\times N$ LDPC matrix and can be used to generate an LDPC code. Now, if we further restrict those permutation matrices to be \textit{circulant} matrices, then the output would be the parity-check matrix for a $b$-QC LDPC code. Unlike standard linear codes, the generator matrix of every QC code cannot be written in a systematic form\footnote{ Note that by systematic form, we particularly mean those encoders whose generator matrices can be written as $[P | I]$, where $I$ is the identity matrix. While it is true that for any linear code one can always find a set of columns that can be used as systematic bits, these positions might not be consecutive.  Since reordering the bits destroys the QC property, it is not possible to put the generator matrix of every QC code into a systematic form mentioned above without violating the QC property.} without violating the QC property. However, many existing QC LDPC ensembles including the ones used for simulation and for the proofs satisfy those constraints.




In order to use the above QC codes for transmission, we modulate the codeword symbols onto elements in a constellation $\mc{A}$ to form the transmitted signals. Throughout the paper, we consider using QC codes over a prime field $\mbb{F}_p$ and restrict the constellation $\mc{A}$ to be pulse amplitude modulation (PAM) with $p$ elements, i.e., $\mc{A}=\{ -\frac{p-1}{2}, \ldots, 0, \ldots, \frac{p-1}{2}  \}$. The mapping $\mc{M}:\mbb{F}_p\rightarrow \mc{A}$ is given by
\begin{equation}\label{eqn:mapping}
    \mc{M}(u)\defeq \left\{ \begin{array}{ll}
    u, & 0\leq u \leq \frac{p-1}{2}, \\
    u-p,  & \frac{p-1}{2}< u < p, \\
    \end{array}\right.
\end{equation}
for $p\geq 3$, and $\mc{M}(u)\defeq u-\frac{1}{2}$ for $p=2$ (i.e., BPSK). This mapping has the important property that $\mc{M}(u \oplus v) = \mc{M}(u) + \mc{M}(v)\hspace{-3pt}\mod p$ and $\mc{M}(u \otimes v) = \mc{M}(u) \cdot \mc{M}(v)\hspace{-3pt}\mod p$ for $p\geq 3$ and $\mc{M}(u \oplus v) +\frac{1}{2}= \mc{M}(u)+\frac{1}{2}+ \mc{M}(v)+\frac{1}{2} \hspace{-3pt}\mod 2$ and $\mc{M}(u \otimes v) +\frac{1}{2}= (\mc{M}(u) +\frac{1}{2})\cdot (\mc{M}(v)+\frac{1}{2})\hspace{-3pt}\mod 2$ for $p=2$.\footnote{Mappings having such properties between two rings are said to be ring homomorphisms. A ring homomorphism is a ring isomorphism if it is bijective. One can see that the mapping in \eqref{eqn:mapping} is in fact a natural mapping from $\mbb{F}_p$ to $\mbb{Z}/p\mbb{Z}$ which is known to be a ring isomorphism. There are other constellations possessing such properties (e.g. \cite{Feng10} \cite{Engin12} \cite{product_const14}) but we restrict ourselves to PAM for brevity.} Note that the above operation is precisely the standard procedure for constructing a lattice from linear codes via Construction A \cite{LeechSloane71} \cite{conway1999sphere}. In fact, one can easily show that the above construction results in lattices having QC property, i.e., QC lattices. Moreover, from a result by Forney \textit{et al.} \cite{forney2000}, we know that applying a capacity-achieving linear code to Construction A would result in a sphere-bound-achieving (or Poltyrev good) lattice. Therefore, existing good QC codes such as AR4JA codes \cite{CCSD} can be adopted to generate good QC lattice codes via Construction A.

\section{Proposed Interleave/Deinterleave Transformed Quasi-Cyclic Code}\label{sec:proposed_framework}
Even though our ultimate application is in networks, in this section, we start with the point-to-point communication to facilitate the illustration of the proposed IDT transform. Consider point-to-point communication with additive white Gaussian noise (AWGN) and with time delays $\tau$ that is upper bounded by the maximal possible delay $D_{max}$. We assume that the transmitter only has access to $D_{max}$ but the receiver knows both $\tau$ and $D_{max}$. For the point-to-point case, one can easily achieve the capacity by using a capacity-achieving code since $\tau$ is known and can be easily compensated. However, in a network where there are multiple source nodes, the signals may arrive at different time and are all mixed together so that this simple approach may no longer work. In order to obtain insight into this problem, we begin with the point-to-point case. Motivated by this issue, we propose a general framework called IDT which utilizes a pair of interleaver/deinterleaver to transform the received signal into the desired form. When combined with QC codes, the IDT allows us to combat time delays that may be introduced in many practical scenarios such as asynchronous channels and/or ISI channels. We nickname this combination as interleave/deinterleave transformed quasi-cyclic (IDT-QC) codes.

\subsection{System Model}
Consider a point-to-point communication with AWGN and delay $\tau\in\{0,\ldots,D_{max}\}$. The transmitter wishes to send a message $\mathbf{w}\in\mbb{F}_p^K$ to the receiver. It first feeds the message into an encoder $\mc{E}^N:\mbb{F}_p^K\rightarrow\mbb{F}_p^N$ to form the codeword $\mathbf{c}=\mc{E}^N(\mathbf{w})\in\mbb{F}_p^N$. The transmitter adopts the modulation scheme $\mc{M}:\mbb{F}_p^N\rightarrow \mc{A}^N$ to form the transmitted signal $\mathbf{x}=\mc{M}(\mathbf{c})\in\mc{A}^N$ where $\mc{A}$ is the signal constellation. The transmitted signal $\mathbf{x}$ is subject to an input power constraint $P$.
\begin{equation}
    \frac{1}{N}\| \mathbf{x}\|^2 = \frac{1}{N}\sum_{n=1}^N |x[n]|^2 \leq P.
\end{equation}
The received signal is then given by
\begin{equation}
    y[n] = x[n-\tau] + z[n], ~~n\in\{1,2,\ldots,N+\tau\},
\end{equation}
where $\tau\in\{0,\ldots,D_{max}\}$ represents an integer delay and $z[n]\sim \mc{N}(0,1)$. Upon receiving $\mathbf{y}$, the receiver then forms an estimate of the message $\hat{\mathbf{w}}\in\mbb{F}_p^K$ via $\mc{G}^N:\mbb{R}^N\rightarrow \mbb{F}_p^K$. Throughout the paper, we assume that $\tau$ is unknown to the transmitter and is known to the receiver and $D_{max}$ is known to both ends. For the ISI channel, this assumption implies that although the transmitter may not know how many taps we would have, it knows the maximal delay spread $D_{max}$. For asynchronous communication, this assumption models the scenario where there is only a very loose synchronization mechanism that would control the time delays to some degree $D_{max}$.
%
In what follows, we give the definition of codes, achievable rates, and capacity.
\begin{define}[Codes]
    An $(N,K)$ code consists of a pair of encoding/decoding functions $(\mc{E}^N, \mc{G}^N)$ described above and an error probability given by
    \begin{equation}
        P_e^{(N)} \triangleq \Pp\left(\hat{\mathbf{w}} \neq \mathbf{w}\right).
    \end{equation}
\end{define}

\begin{define}[Achievable rate and capacity]
    For a given set of parameters $P$ and $D_{max}$, a rate $R(P,D_{max})$ is achievable if for any $\varepsilon>0$ there is an $(N,K)$ code over $\mbb{F}_p$ such that
    \begin{equation}
        K\geq NR(P,D_{max})/\log(p) \text{~and~} P_e^{(N)}\leq \varepsilon.
    \end{equation}
    The capacity is defined as the supremum of all achievable rates given by
    \begin{equation}
        C(P,D_{max})\defeq \sup R(P,D_{max}).
    \end{equation}
\end{define}

\subsection{IDT-QC Codes}
\begin{figure}
    \centering
    \includegraphics[width=3in]{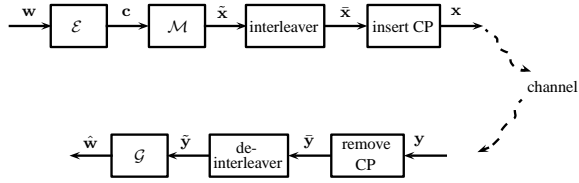}
    \caption{Block diagram of the proposed IDT.}
    \label{fig:IDT}
\end{figure}

\begin{figure}
    \centering
    \includegraphics[width=3in]{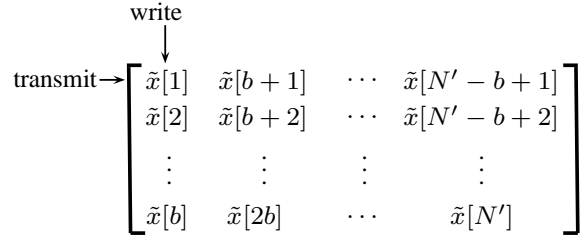}
    \caption{The write column-wise transmit row-wise interleaver.}
    \label{fig:interleaver}
\end{figure}

\begin{figure}
    \centering
    \includegraphics[width=3in]{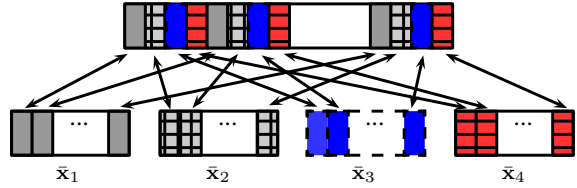}
    \caption{An example of the IDT-QC codes with $b=4$.}
    \label{fig:inter_deinterx}
\end{figure}

Let $\mc{C}$ be a $(N',K)$ $b$-QC linear/lattice code with the design rate $R_d = r_d\cdot\log(p)$ where $r_d\defeq K/N'$ and $r_d b\in\mbb{Z}$. Also, we enforce the generator matrix of this code to be systematic. In the proposed IDT-QC codes shown in Fig.~\ref{fig:IDT}, the transmitter maps the message to a codeword $\mathbf{c}\in\mc{C}$ via the encoder $\mc{E}$. This codeword is modulated by $\mc{M}$ to form the signal $\tilde{\mathbf{x}}$. The signal is then fed into a $(b,N'/b)$ write column-wise transmit row-wise interleaver \cite{Guess00} to get a interleaved signal $\bar{\mathbf{x}}$ where the input-output relationship is given by
\begin{equation}
    \bar{x}[n] = \tilde{x}\left[1+(\lfloor n/L \rfloor)+b\cdot(n\mod L-1)\right],
\end{equation}
where $L \defeq N'/b$ is always an integer provided by the QC constraint. An illustration of interleaving can be found in Fig.~\ref{fig:interleaver} and one example with $b=4$ is given in Fig.~\ref{fig:inter_deinterx}.

\begin{figure}
    \centering
    \includegraphics[width=3in]{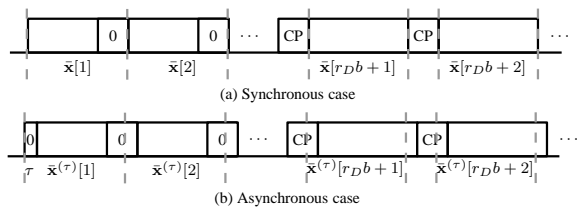}
    \caption{(a) Overall signal structure. (b) Asynchronous case.}
    \label{fig:signal_struct}
\end{figure}

Note that one can write the interleaved codeword as the collection of $b$ sub-blocks as
\begin{equation}
    \bar{\mathbf{x}} = [\bar{\mathbf{x}}[1] \bar{\mathbf{x}}[2] \ldots \bar{\mathbf{x}}[b]],
\end{equation}
where each sub-block $\bar{\mathbf{x}}[s]$ for $s\in\{1,\ldots,b\}$, is of length $L$. For each of the first $r_d b$ sub-blocks, we freeze the $D_{max}$ last positions to be zero. This is possible since the encoder is systematic and the first $r_d b$ blocks correspond to the message part. We then insert a cyclic prefix (CP) of length $D_{max}$ for each of the last $(1-r_d)b$ sub-blocks by appending the last $D_{max}$ symbols to the front. The overall transmitted signal is given by
\begin{equation}
    \mathbf{x} = [ \mathbf{x}[1], \mathbf{x}[2], \ldots, \mathbf{x}[b]],
\end{equation}
where for $s\in\{1,\ldots,r_d b\}$, $\mathbf{x}[s]=\bar{\mathbf{x}}[s]$ whose last $D_{max}$ symbols are 0, and for $s\in\{r_d b+1,\ldots,r_d b\}$
\begin{equation}
    \mathbf{x}[s] \defeq [\underset{\text{CP with length $D_{max}$}}{\underbrace{\bar{x}^{L-D_{max}+1}[s],\ldots,\bar{x}^{L}[s]}}, \underset{= \bar{\mathbf{x}}[s]}{\underbrace{\bar{x}^1[s], \ldots, \bar{x}^L[s]}}].
\end{equation}
The total length of this signal is $N=N'+(1-r_d)b D_{max}$. An illustration of the overall signal structure is given in Fig.~\ref{fig:signal_struct}.(a). In fact, for the purpose of IDT transform, one does not have to distinguish the parts using CPs and freezing symbols; it suffices to append CPs for all the sub-blocks. The reason that we choose to freeze symbols instead of inserting CPs for the first $r_d b$ sub-blocks will become apparent in Sections \ref{sec:IF_ISI} and \ref{sec:asyn_CF}.

\begin{figure}
    \centering
    \includegraphics[width=3in]{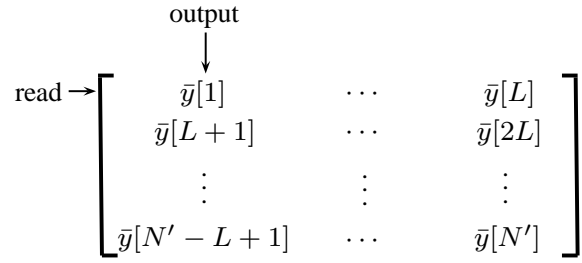}
    \caption{The read row-wise output column-wise deinterleaver.}
    \label{fig:deinterleaver}
\end{figure}

At the receiver end, since the receiver knows the time delays $\tau$ and $\tau\leq D_{max}$, it first discards the CP for each sub-block to form $\bar{\mathbf{y}}$. As shown in Fig.~\ref{fig:deinterleaver}, this signal $\bar{\mathbf{y}}$ is then fed to a $(b,N'/b)$ read row-wise output column-wise deinterleaver to get output $\tilde{\mathbf{y}}$ where the input-output relationship is given by
\begin{equation}
    \tilde{y}[n] = \bar{y}\left[1+(\lfloor n/b \rfloor)+L\cdot(n\mod b-1)\right],
\end{equation}
which is then fed into the decoder of the QC code to form an estimate of the message.
The actual rate of this IDT-QC code is given by
\begin{align}\label{eqn:R_a}
    R_a &= \frac{K-r_d b D_{max}}{N'+(1-r_d)b D_{max}}\log(p) \nonumber \\
    &= \left(1 - \frac{(2-r_d)bD_{max}}{N'+(1-r_d)bD_{max}}\right)R_d
\end{align}
which tends to $R_{d}$ asymptotically but introduces a rate loss for any finite $N'$.

In the following, we provide some properties of the IDT-QC codes.
\begin{lemma}\label{lma:transform}
    If the receiver opts not to compensate the delay $\tau$, i.e., the receiver observes a noisy version of the transmitted signal delayed by $\tau$, $[ \underset{\tau}{\underbrace{0,\ldots,0}}, \mathbf{x}[1], \mathbf{x}[2], \ldots, \mathbf{x}[b]]$,
then the proposed IDT transforms the received signal into a noisy version of the codeword circularly shifted by $b\cdot\tau$. Moreover, if a $b$-QC code is employed in conjunction with IDT, one can directly decode $\mathbf{c}^{(b\tau)}$.
\end{lemma}
\begin{IEEEproof}
    Let us first assume that there is no channel noise. For $\tau\leq D_{max}$, due to the frozen bits for the first $r_d b$ sub-blocks and the insetion/removal of the CPs for the last $(1-r_d)b$ sub-blocks, the linear shift by $\tau$ introduced by the channel has been transformed into circular shift of each sub-block by $\tau$. This is written with a slight abuse of notation as
    \begin{equation}\label{eqn:x_bar_d}
        \bar{\mathbf{x}}^{(\tau)} \defeq [ \bar{\mathbf{x}}^{(\tau)}[1], \bar{\mathbf{x}}^{(\tau)}[2], \ldots, \bar{\mathbf{x}}^{(\tau)}[b]],
    \end{equation}
    where for $s\in\{1,\ldots,b\}$,
    \begin{equation}
        \bar{\mathbf{x}}^{(\tau)}[s] = [\bar{x}^{L-\tau+1}[s],\ldots,\bar{x}^{L}[s],\bar{x}^{1}[s],\ldots,\bar{x}^{L-\tau}[s]]
    \end{equation}
    is the circularly shifted version of $\bar{\mathbf{x}}[s]$ by $\tau$ positions. One can then verify that the output of the deinterleaver with this input would be $\tilde{\mathbf{x}}^{(b\tau)}$ which is corresponding to the codeword $\mathbf{c}^{(b\tau)}$ if a $b$-QC code is employed. Therefore, in the presence of channel noise, the received signal would be a noisy signal corresponds to $\mathbf{c}^{(b\tau)}$ and hence we can directly decode $\mathbf{c}^{(b\tau)}$ instead of $\mathbf{c}$.
\end{IEEEproof}

\begin{theorem}\label{thm:IDT-QC_capacity}
    There exists a sequence of IDT-QC linear/lattice codes that achieve the capacity of the asynchronous point-to-point AWGN channel.
\end{theorem}
\begin{IEEEproof}
    Let $L>D_{max}$ and let $\mc{C}^1, \mc{C}^2, \ldots, \mc{C}^L$ be identical $(b,k)$ linear/lattice codes that can approach the capacity  \cite{Gallager} \cite{erez04} when $b\rightarrow\infty$ for a fixed $k/b$. i.e., for a $\varepsilon>0$, there is a large enough $b$ such that $k/b\log(p) >C(P,0)-\varepsilon$ and $P_e^{(b)}<\varepsilon/L$. Moreover, in order to make these codes fit into the aforementioned form, the generator matrices of these codes should be systematic. We would like to construct a capacity-achieving IDT-QC codes $\mc{C}$ for the asynchronous AWGN channel from $\mc{C}^1, \mc{C}^2, \ldots, \mc{C}^L$. For every $\mathbf{c}^l\in \mc{C}^l$ for $l\in\{1,2,\ldots, L\}$, we construct the codeword
    \begin{equation}
        \mathbf{c} = [\mathbf{c}^1,\mathbf{c}^2,\ldots,\mathbf{c}^L].
    \end{equation}
    Using the fact that $\mc{C}^1, \mc{C}^2,\ldots, \mc{C}^L$ are identical linear/lattice codes, one can see that the collection of such codewords forms a $(bL,kL)$ $b$-QC code with design rate $R_d = r_d\log(p)$ where $r_d = K/N' = k/b$. The codeword is then modulated to $\tilde{\mathbf{x}}$, fed into the interleaver to form $\bar{\mathbf{x}}$, bits-frozen and CP-appended to get $\mathbf{x}$.

    The receiver observes a noisy version of the transmitted codeword delayed by $\tau$ which can be easily compensated as $\tau$ is known by the receiver. It then removes the CP and feeds the signal to the deinterleaver to get a noisy version of the original signal $\mathbf{x}$. The error probability of this IDT-QC code can be bounded using the union bound as
    \begin{equation}
        P_e^{(Lb)} < L\cdot \frac{\varepsilon}{L}=\varepsilon,
    \end{equation}
    and from \eqref{eqn:R_a}, one has the actual rate given by
    \begin{equation}
        R_a(b,L)=\left(1-\frac{(2b-k)D_{max}}{bL+(b-k)D_{max}}\right)R_d.
    \end{equation}
    Now, letting $b$ go to infinity results in $R_d\rightarrow C$ and
    \begin{equation}
        R_a(L)=\lim_{b\rightarrow\infty}R_a(b,L)=\left(1-\frac{(2-r_d)D_{max}}{L+(1-r_d)D_{max}}\right)C,
    \end{equation}
    which in turn results in $\lim_{L\rightarrow\infty}R_a(L)=C$ which completes the proof.
\end{IEEEproof}



\begin{remark}
    Note that the ensemble of codes that we construct in Theorem~\ref{thm:IDT-QC_capacity} is not necessarily a good ensemble in the sense that for a particular target error rate, it requires a very long code length to achieve that error rate. In practice, QC codes are usually not constructed this way and QC codes with not very long block length comparing to the one constructed here can perform very well. Since this is the case, we only use this ensemble for the proof and construct QC codes for simulation by existing constructions such as AR4JA.
\end{remark}



\subsection{Advantages and Disadvantages}
 Here, we provide a discussion of some advantages and disadvantages of the proposed IDT-QC codes.

\textit{\textbf{Advantages}}:
\begin{itemize}
    \item The proposed IDT-QC codes substantially generalize the idea of \cite{FuShenli09}. Unlike \cite{FuShenli09} which only considers allowing the use of convolutional codes and only works for the two-way relay channel, the proposed framework is more general in that it can take any QC codes and would work for a larger class of networks as will be shown later on. The use of QC codes provides significant improvement in practice as there are families of QC codes (e.g. AR4JA codes \cite{Abbasfar07}) that can work very close to the Shannon limit with reasonable decoding complexity (iterative decoding). On the other hand, for convolutional codes, one has to use a very large constraint length (usually with formidably high decoding complexity) in order to approach the Shannon limit.

    \item Compared to the use of cyclic codes as in \cite{ordentlich12} \cite{wu13}, our approach enjoys a better error-correcting capability provided by QC codes. This can be easily seen by the fact that QC codes contain the family of cyclic codes as a special case. Our simulation in the following sections show that even when we consider the rate loss introduced by the IDT, IDT-QC codes significantly outperform cyclic codes. Moreover, unlike cyclic codes, the proposed IDT-QC codes can be easily shown to achieve the Shannon limit.

    \item In practice, QC codes (QC LDPC codes in particular) have been very popular and have been adopted in many communication standards due to the existence of efficient encoding and decoding algorithms. The proposed IDT-QC codes naturally inherit those practical benefits from QC codes and hence are practically attractive.

    \item As will be unveiled in the following sections, the proposed IDT-QC codes can not only harness interference in the presence of asynchronism, but also exploit asynchronism in some cases.
\end{itemize}

\textit{\textbf{Disadvantages}}:
\begin{itemize}
    \item Due to the use of interleaver and deinterleaver in our proposed IDT-QC codes, the transmitter has to wait until the entire codeword is generated before transmission. This results in an increased encoding latency.

    \item While the rate loss is negligible as $L\rightarrow\infty$ when we prove the capacity result in Theorem~\ref{thm:IDT-QC_capacity}, it must be taken into account in the finite length regime. However, in the following sections, we will show that the proposed IDT-QC codes outperform cyclic codes even when rate loss is included.
\end{itemize}

\section{Application 1: Integer-Forcing Equalization for ISI Channels}\label{sec:IF_ISI}
%

In this section, we consider point-to-point communication with ISI. It has been known that the capacity of the ISI channel can be achieved by multi-carrier systems. However, the high peak-to-average power ratio makes this approach less attractive for applications requiring extremely low complexity such as wireless sensor networks. Another way to approach capacity is to code over time domain and uses a decision feedback equalizer (DFE) at the receiver. For this to work, a very long interleaver/deinterleaver (between multiple codewords) is required to avoid error propagation. Recently, Ordentlich and Erez in \cite{ordentlich12} proposed a new linear equalization technique called integer-forcing equalization. This new equalization technique does not require inteleaver/deinterleaver between codewords and avoids the error propagation as no DFE is implemented. However, one of the drawbacks pointed out by the authors themselves is that the channel coding adopted is required to be cyclic and hence is not guaranteed to achieve capacity. In what follows, we will replace the cyclic codes by the proposed IDT-QC codes which are capacity-achieving to achieve the upper bound on information rates presented in \cite{ordentlich12}. It should be noted that unlike the DFE-based scheme, the interleaver/deinterleaver for the proposed framework is within a QC codeword and hence is much shorter than that in DFE-based schemes.

\subsection{Problem Statement}
The transmitter encodes its message $\mathbf{w}\in\mbb{F}_p^K$ to a codeword $\mathbf{c}\in\mbb{F}_p^N$ which is then mapped to a signal $\mathbf{x}\in\mc{A}^N$ via $\mc{M}$ where $\mc{A}$ is the signal constellation (e.g., M-PAM) and $\mc{M}$ is the natural mapping. This signal is subject to a power constraint $P$ and is sent over an AWGN channel with ISI $\mathbf{h}=[h_1,\ldots,h_{d_M}]$ where $d_M$ depends on the maximal delay spread and the sampling frequency. The received signal is given by
\begin{equation}
    \mathbf{y} = \mathbf{h} * \mathbf{x} + \mathbf{z}.
\end{equation}
i.e., the received signal would be a noisy version of a linear combination of the codeword linearly shifted by integers. We consider a recently proposed linear equalizer called integer-forcing equalizer proposed in \cite{ordentlich12}. This technique first passes $\mathbf{y}$ to a linear equalizer chosen in such a way that the equalized channel impulse responses are forced to be an integer vectors $\mathbf{i}\defeq [i_0, i_1,\ldots,i_{D_{max}}]$. Also, one can easily transform linear convolution into circular convolution. The equalized signal is then given by
\begin{equation}
    \mathbf{y} = \sum_{d=0}^{D_{max}} i_d \mathbf{x}^{(d)} + \mathbf{z}',
\end{equation}
where $\mathbf{z}'$ is the filtered noise.

The authors in \cite{ordentlich12} then proposed using cyclic codes over $\mbb{F}_p$ at the transmitter so that $\varphi(\sum_{d=0}^{D_{max}} i_d \mathbf{x}^{(d)})$ with $\varphi\defeq\mc{M}^{-1}\circ\mod p$ is a codeword of the same cyclic code. Therefore, one can directly decode $\varphi(\sum_{d=0}^{D_{max}} i_d \mathbf{x}^{(d)})$ from $\mathbf{y}\hspace{-3pt}\mod p$. This decoded signal is then used to recover $\mathbf{x}$ and hence $\mathbf{w}$. In what follows, we propose using the IDT-QC codes to replace the cyclic codes. Since the problem of designing and analyzing integer-forcing equalizers has been well addressed in \cite{ordentlich12}, we assume that the ISI channel has already been integral. i.e., $\mathbf{h}=\mathbf{i}$.

\begin{figure}
    \centering
    \includegraphics[width=3in]{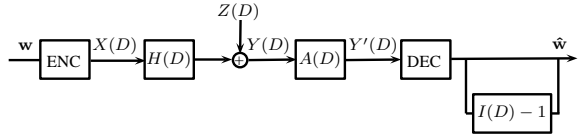}
    \caption{The integer-forcing equalization system.}
    \label{fig:ISI_model}
\end{figure}


\subsection{Using IDT-QC for Point-to-Point Communication with ISI}
\begin{figure}
    \centering
    \includegraphics[width=3in]{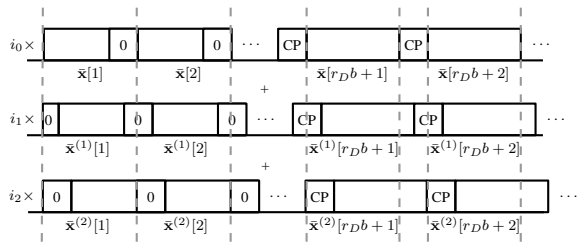}
    \caption{The idea of using IDT-QC for point-to-point communication with ISI.}
    \label{fig:IDT_QC_ISI}
\end{figure}

After the integer-forcing equalizer, the signal becomes linear combination of linearly-shifted versions of the transmitted signal with integer coefficients $i_d\in\mbb{Z}$ for $d\in\{0,\ldots,D_{max}\}$. As shown in Fig.~\ref{fig:IDT_QC_ISI} where an example with $D_{max}=2$ is given, the receiver then removes the received signal at the positions where the CP of the first tap's signal should be. Then the signal can be expressed as a noisy version of the linear combination of the codeword circularly shifted by integers given by
\begin{equation}
    \bar{\mathbf{y}} = \sum_{d=0}^{D_{max}} i_d \bar{\mathbf{x}}^{(d)} + \bar{\mathbf{z}},
\end{equation}
where $\bar{\mathbf{x}}^{(d)}$ is as in \eqref{eqn:x_bar_d} and elements in $\bar{\mathbf{z}}$ and those in $\mathbf{z}$ have the same distribution. As shown in Lemma~\ref{lma:transform}, any linear shift by an integer $d\leq D_{max}$ introduced by the channel will be transformed by the IDT into $b\cdot d$ circular shift for the codeword. i.e., the channel would be transformed into
\begin{equation}\label{eqn:ISI_after_IDT}
    \tilde{\mathbf{y}} = \sum_{d=0}^{D_{max}} i_d \tilde{\mathbf{x}}^{(bd)} +\tilde{\mathbf{z}},
\end{equation}
where elements in $\tilde{\mathbf{z}}$ and those in $\mathbf{z}$ have the same distribution. Moreover, since IDT-QC codes adopt $b$-QC codes for channel coding, every $\tilde{\mathbf{x}}^{(bd)}$ in \eqref{eqn:ISI_after_IDT} corresponds to a valid codeword in the underlying QC code. This in turn allows us to directly decode $\tilde{\mathbf{y}}\hspace{-3pt}\mod p$ to a valid codeword $\varphi\left(\sum_{d=0}^{D_{max}} i_d \tilde{\mathbf{x}}^{(bd)}\right)$ (or $\varphi(I(D^b)X(D))$ in the $D$-domain) in the same QC code. After this codeword is decoded, one can use the knowledge of frozen bits to strip out all the message bits. This is perhaps easier seen from the second representation and is shown in Fig.~\ref{fig:IDT_QC_ISI}. It can be seen that within each sub-block corresponding to the message part, one can initiate the deconvolution since the last $D_{max}$ bits are frozen.

It has been shown in Section~\ref{sec:proposed_framework} that there exists a sequence of the proposed IDT-QC codes that can achieve capacity. Thus, theoretically, using the proposed framework allows one to achieve the upper bound on information rates presented in \cite{ordentlich12} which may not be achievable for the cyclic coding scheme proposed therein. Therefore, the proposed framework bridges the gap-to-capacity for such integer-forcing equalization schemes. In what follows, we provide some simulation results to demonstrate that the proposed IDT-QC codes outperform cyclic codes even though for the finite-length regime, the proposed IDT-QC codes suffer from a rate loss. It is worth mentioning that in addition to being of independent interest, the integer-forcing equalization for ISI channel will play an important role in using IDT-QC for asynchronous compute-and-forward.

\subsection{Simulation Results}
\begin{figure}
    \centering
    \includegraphics[width=3in]{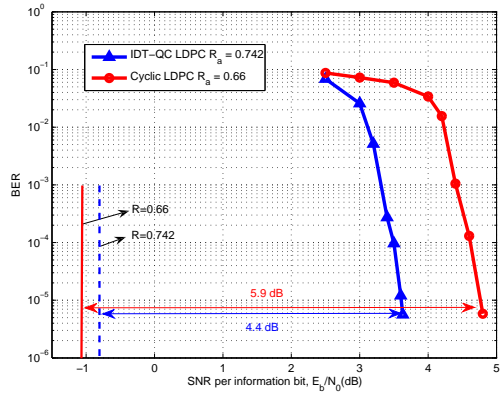}
    \caption{BER comparison of the IDT-QC LDPC code with $R_a=0.742$ and the cyclic LDPC code with $R_a=0.66$ over the dicode channel $(1+D)$. The dashed and solid vertical lines are the required SNRs corresponding to the information rates $R=0.742$ and $R=0.66$, respectively.}
    \label{fig:ISI-ber}
\end{figure}

We now provide some simulation results to compare the proposed IDT-QC framework and the cyclic coded scheme proposed in \cite{ordentlich12}. We consider the dicode channel whose impulse response is $I(D)=1+D$ (i.e., $i_0=i_1=1$); therefore, $D_{max}=1$. We construct an binary IDT-QC LDPC code from the AR4JA ensemble \cite{CCSD} with $N' = 4096$, $b=32$, $K=3072$, and the design rate $R_d = 0.75$. The actual rate of this code is $R_a = 0.742$. For comparison with the scheme in \cite{ordentlich12}, we also construct a cyclic LDPC code from the ensemble proposed in \cite{Huang12} with $N' = N = 4095$, $K=2703$, and the design rate $R_d = 0.66$. Note that for the cyclic coded scheme in \cite{ordentlich12}, one has to freeze $D_{max}$ bits for initializing the deconvolution. This results in the actual rate $R_a \approx R_d = 0.66$. For the both codes, the decoding algorithm is a message-passing algorithm \cite{ModernCodingTheory} with at most $200$ iterations. Since $I(D)=1+D$, the receiver attempts to decode $\mathbf{c} \oplus \mathbf{c}^{(b)}$. Simulation results presented in Fig.~\ref{fig:ISI-ber} show that in spite of having a higher rate, the proposed IDT-QC LDPC code provides roughly 1.1 dB gain when BER is at $10^{-5}$. This is mainly because the proposed IDT transform enables the use of QC codes where very powerful ensembles such as AR4JA can be easily constructed. Moreover, the conventional scheme in \cite{ordentlich12} relies solely on the family of cyclic codes which is much smaller than that of QC codes.

 We also provide the information rate corresponding to the independent uniformly distributed input distribution for the dicode channel estimated by the method in \cite{pfister01} (which is equivalent to the forward recursion of the BCJR algorithm). One observes that there is a roughly 4.4dB gap between the proposed scheme and the corresponding information rate at $P_e \approx 10^{-5}$. This gap comes from the following sources. First and foremost, a receive filter has not been used to harness all the energy in all the taps of the ISI channel. In this example, this contributes a 3~dB loss. A second source is the power loss inherited from the integer-forcing equalization approach which transforms the channel into a $\hspace{-3pt}\mod p$ channel (here $\hspace{-3pt}\mod 2$). The final source of this gap simply comes from the fact that the block length we consider here (4096) is rather small. Similar but larger gap can be observed for the cyclic coded integer-forcing equalization scheme.

It should be noted that for a single user ISI channel, using conventional equalization techniques such as a decision feedback equalizer (DFE) will provide better results than integer-forcing equalization. However, when a compute-and-forward problem is considered with multiple users and ISI, conventional equalization techniques will not be sufficient to efficiently compute functions of transmitted signals since the interference from multiple users and ISI cannot be simultaneously removed easily. In these cases, integer-forcing equalization can substantially outperform conventional equalization techniques.

\section{Application 2: Asynchronous Compute-and-Forward}\label{sec:asyn_CF}
In this section, we study the compute-and-forward relay network introduced by Nazer and Gastpar \cite{nazer2011CF}. In particular, we consider the asynchronous version of this network where signals sent from different source nodes may arrive at a destination node at different times. In the synchronous case, the compute-and-forward strategy suggested in \cite{nazer2011CF} implements an \textit{identical} nested lattice code \cite{erez04} at each user and directly decodes the received signal to a modulo version of linear combination of the codewords with integer coefficients at the destination. This scheme is shown to provide a substantially higher computation rate in the medium signal-to-noise ratio (SNR) regime than existing schemes. For practical purposes, in \cite{Engin12}, this nested lattice code is replaced by linear code over $\mbb{F}_p$ together with a signal mapping possessing the property described in Section~\ref{sec:prelim} in order to exploit the structural gain. The destination then decodes the received signal to a linear combination of the codewords over $\mbb{F}_p$.

There have been a few attempts at using physical-layer network coding to this asynchronous setting. A convolutional coded scheme has been proposed in \cite{FuShenli09} to deal with synchronization errors; however, only integer-valued delays (i.e., frame-level asynchronism) are allowed. In \cite{Lu12}, an over-sampling method was proposed and a graph-based decoding algorithm has been proposed specifically for this over-sampling model. This over-sampling method can take real-valued delays (i.e., symbol-level asynchronism) but only within one symbol time; thus, results in a stringent timing synchronization. In \cite{najafi13}, frame-level and symbol-level asynchronous compute-and-forward are considered where the destinations are only able to compute synchronous functions. A very recent work in \cite{wu13} has successfully applied cyclic codes to this problem so that asynchronous functions are computable and showed through simulation that cyclic codes are able to combat with real-valued delays within one \textit{packet} time.

In this section, we replace the cyclic codes by the proposed IDT-QC codes and show that this replacement allows us to prove capacity results for both the frame-level and symbol-level cases. Moreover, the simulation results given in Section~\ref{sec:practical} show that this replacement substantially improves the performance. It should be noted that since the proposed scheme relies on the quasi-cyclic property instead of the cyclic property to deal with synchronization errors, the delay constraint is more stringent than the scheme in \cite{wu13}. Nonetheless, it only requires the delays to be controlled within a certain range $D_{max}$ (say few symbols time), which is practically reasonable.



\subsection{Problem Statement}
\begin{figure}
    \centering
    \includegraphics[width=2.5in]{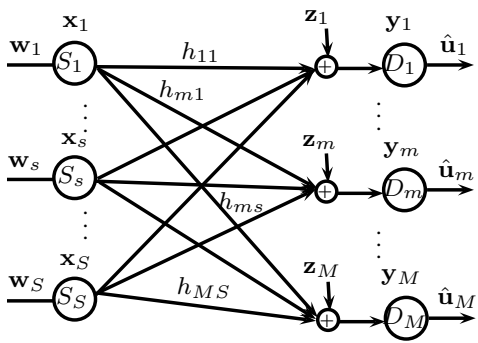}
    \caption{The compute-and-forward relay network.}
    \label{fig:CF_model}
\end{figure}

As shown in Fig.~\ref{fig:CF_model}, in a compute-and-forward network, there are total $S$ source nodes and $M\geq S$ destination nodes. Each source node $s\in\{1,\ldots,S\}$ encodes its message $\mathbf{w}_s\in\mbb{F}_p^K$ to a codeword $\mathbf{c}_s\in\mbb{F}_p^N$. This codeword is then modulated to the transmitted signal $\mathbf{x}_s\in\mc{A}^N$ via a mapping $\mc{M}$ as described in Section~\ref{sec:prelim}.
The codeword is subject to a power constraint given by
\begin{equation}
    \frac{1}{N}\| \mathbf{x}_s\|^2 = \frac{1}{N}\sum_{n=1}^N |x_s[n]|^2 \leq P.
\end{equation}

Let $\tau_{ms}$ be the delay experienced by the signal from source $s$ to destination $m$. We will separately consider two cases, namely the frame-level asynchronous compute-and-forward where $\tau_{ms}\in\{0,\ldots,D_{max}\}$ and the symbol-level asynchronous compute-and-forward where $\tau_{ms}\in[0,T)$ with $T$ being symbol duration. For the frame-level asynchronous one, the received signal is given by
\begin{equation}
    y_m[n] = \sum_{s=1}^S h_{ms}x_s[n-\tau_{ms}] + z_m[n],
\end{equation}
where $h_{ms}\in\mbb{R}$ (or $\mbb{C}$ depending on whether the signal constellation is real or complex) is the channel coefficient between the source node $s$ and destination $m$, and $z_m[n]\sim \mathcal{CN}(0,1)$. For the symbol-level asynchronous compute-and-forward, one has to work with the continuous-time model given by
\begin{equation}\label{eqn:continous_time}
    y_m(t)=\sum_{s=1}^S\sum_{n=1}^{N}h_{ms} x_{s}[n]p(t-nT-\tau_{ms})+z(t),
\end{equation}
where $p(t)$ is the pulse shaping function and $z(t)$ is a Gaussian process with zero mean and variance $1$.

The destination node $m$ is only interested in computing and forwarding a function of the messages. In particular, the compute-and-forward scheme in \cite{nazer2011CF} confines itself to synchronous functions of the messages which mimics the behavior of linear network coding. i.e., the destination node $m$ chooses $\{b_{ms}\}$ and computes $\mathbf{u}_m = \oplus_{s=1}^{L} b_{ms}\mathbf{w}_s$ such that the computation rate at node $m$ is maximized. The computed functions together with $\{b_{ms}\}$ are then forwarded to a central destination which desires all the messages. It is clear that as long as the coefficients $\{b_{ms}\}$ form a full-rank matrix, the central destination would be able to invert the matrix and obtain all the messages. In the sequel, we will show that the use of IDT-QC codes allows one to compute asynchronous functions which may lead one to an increased computation rate. For ease of exposition, we will separately discuss the frame-level and symbol-level models and restrict ourselves to $S=M=2$, but the proposed scheme works for general scenarios.


\subsection{Using IDT-QC for Frame-Level Asynchronous Compute-and-Forward}
\begin{figure}
    \centering
    \includegraphics[width=3in]{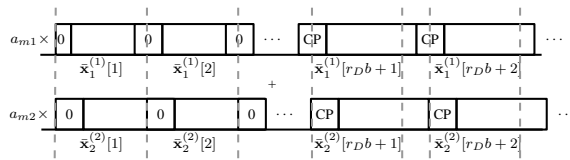}
    \caption{An example of using IDT-QC for asynchronous compute-and-forward where $\tau_{m1}=1$, $\tau_{m2}=2$, and $D_{max}=2$.}
    \label{fig:IDT_QC_CF}
\end{figure}

We illustrate the idea of using the proposed IDT-QC codes for frame-level asynchronous compute-and-forward. Each source node adopts a same $b$-QC code over $\mbb{F}_p$ for encoding its message to the codeword $\mathbf{c}_s$ which is then modulated to the signal $\tilde{\mathbf{x}}_s=\mc{M}(\mathbf{c}_s)$. It will then be interleaved to form $\bar{\mathbf{x}}_s$ and further added frozen bits and appended CPs to form the transmitted signal $\mathbf{x}_s$. The length of the CPs is again set to be $D_{max}$. One difference here is that for a compute-and-forward network with $S$ source nodes, one has to freeze $SD_{max}$ positions instead of $D_{max}$ positions for each sub-block corresponding to message part. As shown in Fig.~\ref{fig:IDT_QC_CF}, the receiver removes the signal at the positions where the first source node's CP should be and the received signal becomes
\begin{equation}
    \bar{\mathbf{y}}_m=a_{m1}\bar{\mathbf{x}}_1^{(\tau_{m1})}+a_{m2}\bar{\mathbf{x}}_2^{(\tau_m2)}+\bar{\mathbf{z}}_{eq,m},
\end{equation}
where $\bar{\mathbf{x}}_s^{(\tau_{ms})}$ is as in \eqref{eqn:x_bar_d} and $\bar{\mathbf{z}}_{eq,m}$ is the effective noise which consists of the noise and the self-interference \cite{nazer2011CF}. The destination node then feeds this signal into the deinterleaver. As discussed in Lemma~\ref{lma:transform}, the proposed IDT-QC codes transform any integer delay $\tau$ introduced by the channel into $b \cdot \tau_{ms}$ circular shifts for the codeword. Thus, the deinterleaver output is given by
\begin{equation}\label{eqn:y_m_tilde}
    \tilde{\mathbf{y}}_m = a_{m1}\tilde{\mathbf{x}}_1^{(b\tau_{m1})} + a_{m2}\tilde{\mathbf{x}}_2^{(b\tau_{m2})} + \tilde{\mathbf{z}}_{eq,m},
\end{equation}
where elements in $\tilde{\mathbf{z}}_{eq,m}$ and that in $\bar{\mathbf{z}}_{eq,m}$ have the same distribution. The receiver $m$ then attempts to compute the lattice point $a_{m1}\tilde{\mathbf{x}}_1^{(b\tau_{m1})} + a_{m2}\tilde{\mathbf{x}}_2^{(b\tau_{m2})}$ and uses the ring homomorphism $\varphi\defeq \mc{M}^{-1}\circ\mod p$ to map this lattice point to
\begin{equation}
    \varphi(a_{m1}\tilde{\mathbf{x}}_1^{(b\tau_{m1})} + a_{m2}\tilde{\mathbf{x}}_2^{(b\tau_{m2})}) =b_{m1}\odot\mathbf{c}_1^{(b\tau_{m1})} \oplus b_{m2}\odot\mathbf{c}_2^{(b\tau_{m2})},
\end{equation}
where $b_{ms}\defeq \varphi(a_{ms})$. It should be noted that since the underlying code we adopt is a $b$-QC code, $\mathbf{c}_s^{(b\tau_{ms})}$ is a codeword and so is $\mathbf{f}_m \defeq b_{m1}\odot\mathbf{c}_1^{(b\tau_{m1})} \oplus b_{m2}\odot\mathbf{c}_2^{(b\tau_{m2})}$.

The computed functions and those coefficients are then forwarded to the central destination and are then further processed to recover all the messages. We now show that the compute-and-forward problem with full rank coefficients can be equivalently represented as ISI channel problems in Section~\ref{sec:IF_ISI} and can be solved by deconvolution if sufficient initial conditions are provided. For the sake of simplicity, we look at the interleaved version of $\mathbf{f}_m$ given by
\begin{equation}
    \bar{\mathbf{f}}_m \defeq b_{m1}\odot\bar{\mathbf{c}}_1^{(\tau_{m1})} \oplus b_{m2}\odot\bar{\mathbf{c}}_2^{(\tau_{m2})}.
\end{equation}
In the $D$-domain, one has that
\begin{align}
    \bar{\mathbf{F}}&=\left(
      \begin{array}{c}
        \bar{F}_1(D) \\
        \bar{F}_2(D) \\
      \end{array}
    \right) \nonumber \\
    &=
    \left(
      \begin{array}{cc}
        b_{11}D^{\tau_{11}} & b_{12}D^{\tau_{12}} \\
        b_{21}D^{\tau_{21}} & b_{22}D^{\tau_{22}} \\
      \end{array}
    \right)
    \odot
    \left(
      \begin{array}{c}
        \bar{C}_1(D) \\
        \bar{C}_2(D) \\
      \end{array}
    \right) \nonumber \\
    &\defeq
    \bar{\mathbf{B}}\odot \bar{\mathbf{C}}.
\end{align}
Note that mathematically, one can now left-multiply by the inverse of the matrix $\bar{\mathbf{B}}$ to get $\bar{\mathbf{C}}$. In order to endow this inverse an operational meaning, we note that for every full rank matrix $\bar{\mathbf{B}}$, one has \begin{equation}
    \bar{\mathbf{B}}^{-1}= \frac{\mathrm{adj}(\bar{\mathbf{B}})}{\det(\bar{\mathbf{B}})},
\end{equation}
where $\det(.)$ is the determinant and $\mathrm{adj}(.)$ is the adjugate. One can then left multiply $\mathbf{F}$ with $\mathrm{adj}(\bar{\mathbf{B}})$ to form
\begin{equation}
    \mathrm{adj}(\bar{\mathbf{B}})\odot \bar{\mathbf{F}} = \det(\bar{\mathbf{B}}) \odot \bar{\mathbf{C}}.
\end{equation}
One observes that the problem has been converted into two separate ISI channel problems whose impulse responses are integer vectors. Moreover, since each element in $\bar{\mathbf{B}}$ has the range $\{0,\ldots,D_{max}\}$, each element in $\det(\bar{\mathbf{B}})$ has range $\{0,\ldots,2D_{max}\}$ or in general $\{0,\ldots,SD_{max}\}$. Therefore, this problem can be solved by deconvolution provided that the transmitter freeze $SD_{max}$ positions for each sub-block belonging to the message part. The actual rate then becomes
\begin{align}\label{eqn:R_a_CF}
    R_a &= \frac{K-r_d b S D_{max}}{N'+(1-r_d)b D_{max}}\log(p) \nonumber \\
    &= \left(1 -\frac{(S + 1 -r_d)D_{max}}{L+(1-r_d)D_{max}}\right)R_d,
\end{align}
which does not affect the asymptotic results. One example is given in the following.
\begin{example}
    Consider a 2-by-2 example over $\mbb{F}_2$. Suppose that relay 1 receives
     \begin{equation}
     y_1[n]=x_1[n-1]+x_2[n]+z_1[n],
     \end{equation}
     and relay 2 receives
     \begin{equation}
     y_2[n]=x_1[n]+x_2[n-1]+z_2[n].
     \end{equation}
     Then
    \begin{equation}\label{eqn:A_ex}
        \bar{\mathbf{B}} = \left(
          \begin{array}{cc}
            D & 1 \\
            1 & D \\
      \end{array}
    \right).
    \end{equation}
    We have $\det(\bar{\mathbf{B}}) = 1+D^2,$ and $\mathrm{adj}(\bar{\mathbf{B}})=\bar{\mathbf{B}}$.
    Thus, by left-multiplying $\mathrm{adj}(\bar{\mathbf{B}})$, one has
    \begin{equation}
        \left(
          \begin{array}{c}
            D\bar{F}_1(D) + \bar{F}_2(D) \\
            \bar{F}_1(D) + D\bar{F}_2(D) \\
          \end{array}
        \right)
        =
        \left(
          \begin{array}{c}
            (1 + D^2)\bar{C}_1(D) \\
            (1 + D^2)\bar{C}_2(D) \\
          \end{array}
        \right),
        \label{eqn:37}
    \end{equation}
    which are two separate ISI channel problems. What (\ref{eqn:37}) implies is that the receiver separately decodes $\bar{\mathbf{c}}_1 \oplus \bar{\mathbf{c}}_1^{(2b)}$ and $\bar{\mathbf{c}}_2 \oplus \bar{\mathbf{c}}_2^{(2b)}$ where $b$ is the shifting constraint. Then using the frozen bits, deconvolutions are performed to obtain $\bar{\mathbf{c}}_1$ and $\bar{\mathbf{c}}_2$
\end{example}

We now present the main theorem of this section.
\begin{theorem}\label{thm:comp_rate_frame}
    Consider the frame asynchronous case where $\tau_{ms}\in\{0,\ldots,D_{max}\}$. At relay $m$, given $\mathbf{h}_m$ and $\mathbf{a}_m$, a computation rate of
    \begin{align}\label{eqn:comp_rate_frame}
        \mathcal{R}(\mathbf{h}_m,\mathbf{a}_m)=\frac{1}{2}\log^{+}\left(\left(\|\mathbf{a}_m\|^{2}-
        \frac{P|\mathbf{h}_m^{H}\mathbf{a}_m|^2}{1+P\|\mathbf{h}_m\|^2}\right)^{-1}\right),
    \end{align}
    where $a_{ms}\in\mbb{Z}$, is achievable per real dimension.
\end{theorem}
\begin{IEEEproof}
    See Appendix~\ref{apx:proof_comp_rate}.
\end{IEEEproof}

\subsection{Achieving higher rates than in the synchronous case}
\label{sec:higher}
    One important observation here is that the proposed QC-IDT scheme allows one to exploit another dimension, namely the delay dimension. This is due to the fact that the QC nature of the proposed scheme enables the computation of asynchronous functions in addition to synchronous ones. Sometimes, this allows one to achieve rates \textit{surpassing} that achieved by tightly synchronous compute-and-forward with the same channel coefficients. For example, the matrix $\bar{\mathbf{B}}_{sync}=\left(
                     \begin{array}{cc}
                       1 & 1 \\
                       1 & 1 \\
                     \end{array}
                   \right)$
    is not invertible; however, the matrix in \eqref{eqn:A_ex} is invertible. It must be noted that the delays are completely determined by the channel so that one does not have control over those parameters. But instead of being limited by those delays, the proposed scheme is capable of exploiting them. One example of how the delay may improve the system performance is given in the following.

\begin{example}
    In Fig.~\ref{fig:async_rate}, we plot the achievable computation rates of asynchronous compute-and-forward, for the cases where $S=M=2$ and $S=M=3$ respectively. The channel coefficients $h_{ms}$ are drawn from i.i.d. Rayleigh distribution. One can see from this figure that for both cases, increasing $D_{max}$ substantially increases achievable computation rates. This effect is most pronounced when $D_{max}$ is increased from 0 to 1. This example demonstrates that when the channel introduces delays, using the proposed scheme which allows the decoding of asynchronous functions results in higher achievable computation rates.


    \begin{figure}
    \centering
    \includegraphics[width=3in]{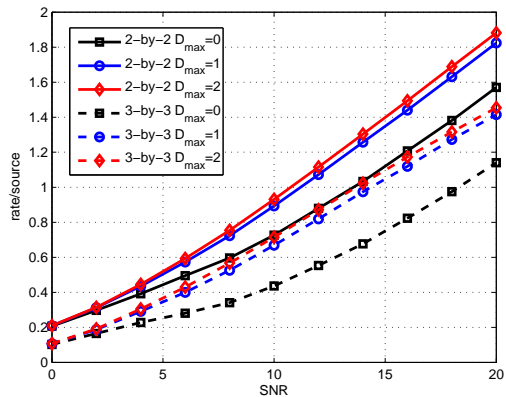}
    \caption{Achievable rates of asynchronous compute-and-forward (average over 10000 realizations).}
    \label{fig:async_rate}
    \end{figure}
\end{example}

\begin{remark}
     Frame asynchronous compute-and-forward has been considered in \cite{najafi13}. The scheme therein does not possess the QC or cyclic properties so that the relays are forced to compute synchronous functions only. As a consequence, they have to use multiple antenna at the relays to rotate the received signal in order to recover synchronous functions in the presence of frame-synchronization errors. This introduces a huge loss not just in rates but also in degrees of freedom because multiple antennas are used just for computing one function at each relay. On the other hand, thanks to the QC nature of the proposed scheme, the computation rates given in Theorem~\ref{thm:comp_rate_frame} have the exactly same form with that in \cite{nazer2011CF}, i.e., as there was no frame asynchronism at all. This is a direct consequence of enabling computing asynchronous functions which are undecodable in \cite{najafi13}. However, this gain does not come for free; this gain comes with an increased burden in the next phase since in addition to $a_{ms}$ (or $b_{ms}$ equivalently), the relays also have to forward the delay profile to the central destination. But this is usually not an issue as the bottleneck is usually the first phase. 
\end{remark}

\subsection{Using IDT-QC for Symbol-Level Asynchronous Compute-and-Forward}
\begin{figure}
    \centering
    \includegraphics[width=3in]{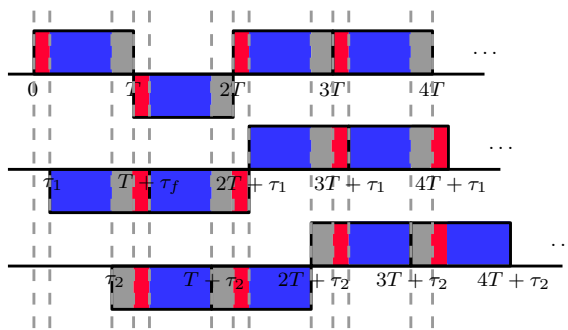}
    \caption{An example of SNR loss resulting from symbol-level asynchronous model where $\tau_{j_1}=0$, $\tau_{j_2}=\tau_1$, and $\tau_{j_2}=\tau_2$.}
    \label{fig:snr_loss}
\end{figure}

We now focus on the symbol-level asynchronous compute-and-forward. i.e., $\tau_{ms}\in[0,T)$ and the continuous-time model in \eqref{eqn:continous_time} is considered. Similar to \cite{Lu12} \cite{YangLiew13} \cite{Zhang09}, we further assume that the pulse shaping function adopted is the ideal (rectangular) pulse.
Let $\pi_m$ be the permutation operation at the relay $m$ defined by
\begin{equation}
    \pi_m(1,\ldots,S) = (j_1,\ldots,j_S),
\end{equation}
such that $\tau_{mj_1}\leq\ldots\leq \tau_{mj_S}$. In order to extract out all the energy, in general, one can perform $S$ different matched filter $p_{mi}(t)$ for $i\in\{1,\ldots,S\}$ as
\begin{equation}\label{eqn:match_filter}
    p_{mi}(t)=\left\{
    \begin{array}{rl}
    0,                                           & {\tau_{mj_{i-1}}+(n-1)T} \leq t < \tau_{mj_i} + (n-1)T \\
    \sqrt{P},                                    & {\tau_{mj_i} + (n-1)T \leq t < \tau_{mj_{i+1}} + (n-1)T}\\
    0,                                           & {\tau_{mj_{i+1}} + (n-1)T \leq t < nT},\\
    \end{array} \right.
\end{equation}
where $\tau_{mj_0}=0$ and $\tau_{mj_{S+1}}=T$ for each $m$. Note that the sampled output of different matched filters would correspond to different functions. For example, as shown in Fig.~\ref{fig:snr_loss}, for the $S=3$ case, three different matched filters would correspond to three different functions, namely, $\mathbf{c}_{j_1}\oplus\mathbf{c}^{(b)}_{j_2}\oplus\mathbf{c}^{(b)}_{j_3}$, $\mathbf{c}_{j_1}\oplus\mathbf{c}_{j_2}\oplus\mathbf{c}^{(b)}_{j_3}$, and $\mathbf{c}_{j_1}\oplus\mathbf{c}_{j_2}\oplus\mathbf{c}_{j_3}$. The corresponding SNR are then given by
\begin{equation}
    P_{mi} = PT(\tau_{mj_i}-\tau_{mj_{i-1}}).
\end{equation}
We then pick the one with the highest SNR for compute-and-forward and discard the others. This would result in the following achievable computation rates.
\begin{theorem}\label{thm:comp_rate_sym}
    Consider the symbol asynchronous case where $\tau_{ms}\in[0,T)$. At relay $m$, given $\mathbf{h}_m$ and $\mathbf{a}_m$, a computation rate of
    \begin{align}\label{eqn:comp_rate_sym}
        \mathcal{R}(\mathbf{h}_m,\mathbf{a}_m)=\frac{1}{2}\log^{+}\left(\left(\|\mathbf{a}_m\|^{2}-
        \frac{P_m|\mathbf{h}_m^{H}\mathbf{a}_m|^2}{1+P_m\|\mathbf{h}_m\|^2}\right)^{-1}\right),
    \end{align}
    where $a_{ms}\in\mbb{Z}$ and $P_m=\max_{i\in\{1,\ldots,S\}} P_{mi}$, is achievable per real dimension.
\end{theorem}
\begin{IEEEproof}
    Similar to the proof of Theorem~\ref{thm:comp_rate_frame} but replacing $P$ by $P_m$.
\end{IEEEproof}

\begin{remark}
    In \cite{najafi13}, the same problem has been studied and achievable computation rates similar to \eqref{eqn:comp_rate_sym} but with $P_m$ replaced by $P_{mj_S}$ has been achieved as only synchronous functions are computable. Our proposed scheme provides increased computation rates through allowing the computation of asynchronous functions. 
\end{remark}

\begin{remark}
    The above scheme only works with the signals corresponding to the function with the highest SNR and completely ignores those corresponding to other functions. However, as will be discussed in the next section, in terms of error probability, one can take advantage of those information by jointly considering detection and decoding.
\end{remark}

\section{Practical Detection and Decoding for Asynchronous Compute-and-Forward}\label{sec:practical}
In this section, we introduce a joint detection and decoding scheme for the proposed IDT-QC codes to alleviate the SNR loss in the presence of symbol-level asynchronism. This decoder is then used for generating simulation results which demonstrate that the proposed framework substantially outperforms the cyclic coding scheme \cite{wu13}. We again begin with the continuous-time model in \eqref{eqn:continous_time}. We further restrict our attention to a specific relay and drop the subscript $m$ for the sake of simplicity. This allows us to assume $\tau_1=0$ and $\tau_2=\tau$ without loss of generality. Let $\tau\in[0,D_{max}]$ where $\tau=\tau_f+\tau_s$ with $\tau_f \in \{0,1,...,D_{max}-1\}$ being the frame-level asynchronism and $\tau_s \in [0,1)$ being the symbol-level asynchronism.

We use a set of matched filters similar to \eqref{eqn:match_filter} to over-sample the received signal \cite{Lu12}. This will result in the following sampled outputs
\begin{equation}
   r[2n-1]= h_1 x_1[n] + h_2 x_2[n-1]+z[2n-1],
\end{equation}
with $x_2[0]=0$ and
\begin{equation}
   r[2n]= h_1 x_1[n] + h_2 x_2[n]+z[2n],
\end{equation}
and $r[2(N-\tau_f)+1]= h_2 x_2[N-\tau_f]+z[2(N-\tau_f)+1]$ where $z[2n-1]\sim \mc{N}(0,1 / \tau_s)$, $z[2(N-\tau_f)+1]\sim \mc{N}(0, 1 / \tau_s)$, and $z[2n]\sim \mc{N}(0,1 / (1-\tau_s))$. In what follows, we describe how to perform the detection and decoding based on this over-sampling model.





\subsection{Joint MAP detection and JCF decoding}

\begin{figure}
    \centering
    \includegraphics[width=3in]{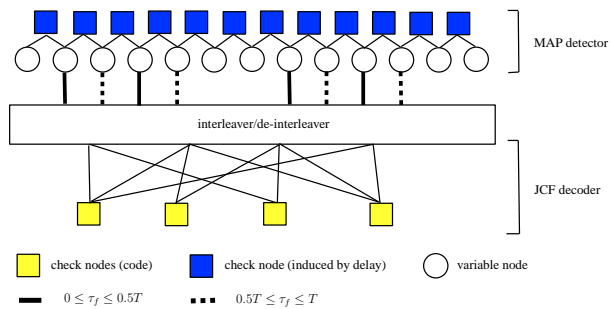}
    \caption{Graph representation of iterative receiver}
    \label{fig:big_graph}
\end{figure}

We now propose a joint detection and decoding scheme which can be deemed as the decoding scheme in \cite{Lu12} tailored specifically for the IDT-QC codes. The decoding algorithm is based on the Tanner graph given in Fig.~\ref{fig:big_graph}. The top part of the Tanner graph with zigzag fashion is associated with the MAP detection which accommodates the correlation between two consecutive over-sampling symbols. The bottom part of the Tanner graph is precisely that of the underlying QC code but over $\mbb{F}_{p^2}$, i.e., the ACNC decoder in \cite{Zhang09} which we refer to as the joint compute-and-forward (JCF) decoder. Unlike \cite{Lu12}, there is a pair of interleaver/deinterleaver between the MAP detection and the JCF decoding parts. Moreover, thanks to the ability described in Lemma~\ref{lma:transform}, depending on the corresponding SNRs, the receiver can opt to decode either $\mathbf{c}_1\oplus \mathbf{c}_2^{(b\tau_f)}$ or $\mathbf{c}_1\oplus \mathbf{c}_2^{(b(\tau_f+1))}$. This is represented as solid and dash edges, respectively.



\subsection{Simulation Results}
We now provide some simulation results. For the sake of simplicity, we only consider coding over $\mbb{F}_2$ with BPSK. The channel parameters are set to be $h_1=h_2=1$ and $D_{max}=5$. We again construct a IDT-QC LDPC code from the AR4JA ensemble with $N=4096$, $b=32$, $K=3072$, and the design rate $R_d=0.75$. The actual rate is then $R_a=0.685$. We would like to recall that for a same set of $D_{max}$ and $b$, the longer the code the smaller the rate loss. Also, we construct the same cyclic LDPC code as in Section~\ref{sec:IF_ISI} with design rate $R_d=0.66$. Note that for using cyclic codes for asynchronous compute-and-forward, one has to freeze $SD_{max}$ bits. Thus, the actual rate of the cyclic LDPC code is $R_a=0.658$. The decoding algorithm for the both codes is the joint MAP detection and JCF decoding described above with $40$ outer iterations and $5$ inner iterations.

\begin{figure}
    \centering
    \includegraphics[width=3in]{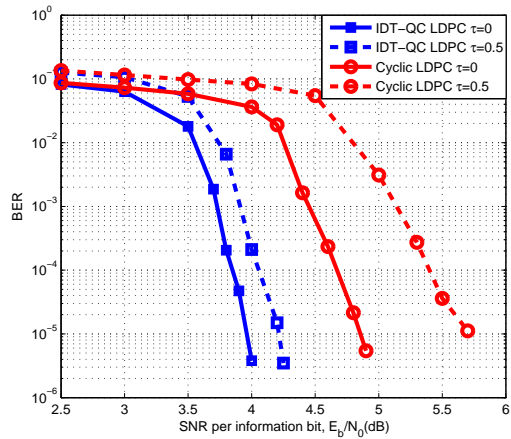}
    \caption{BER performance of the proposed IDT-QC LDPC code with $R_a=0.685$ and the cyclic LDPC code with $R=0.658$ for compute-and-forward.}
    \label{fig:TWR-ber}
\end{figure}

In Fig.~\ref{fig:TWR-ber}, BER versus SNR curve is plotted. One can observe that despite of having a higher rate, the proposed IDT-QC LDPC code outperforms the cyclic LDPC code by roughly 1.1 dB when $\tau=0$, i.e., under perfect synchronization. This is the coding gain offered by the AR4JA code over the cyclic code adopted. When $\tau=0.5$, the proposed IDT-QC LDPC code provides roughly 1.5 dB gain over the cyclic-LDPC considered. This enlarged gap may be explained by the observation that the joint graph of the zigzag detection and the parity check of a cyclic code is more likely to create short cycles compared to QC LDPC codes.

\begin{figure}
    \centering
    \includegraphics[width=3in]{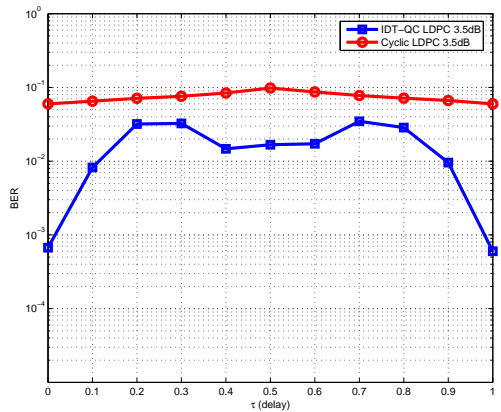}
    \caption{BER versus $\tau$ for the proposed IDT-QC LDPC code with $R=0.685$ and the cyclic LDPC code with $R=0.658$ for compute-and-forward.}
    \label{fig:TWR-ber-tau}
\end{figure}

In Fig. \ref{fig:TWR-ber-tau}, BER versus delay curve is plotted for SNR$=3.5$ dB. One observes that in the region $\tau\in[0,T]$, the proposed IDT-QC LDPC code always performs better than the cyclic LDPC code in terms of BER. Moreover, we observe a symmetric behavior of BER about $\tau=0.5$. This is a consequence of allowing decoding to asynchronous functions so that the performance would only depend on how close the delay $\tau$ is to an integer. According to this observation, one expects a periodic behavior for other $[kT, (k+1)T]$ within $[0,D_{max}]$. Another interesting observation is that there is a local minimum at around $\tau=0.5$. This may be explained by the observation that at around $\tau=0.5$, two codewords are well separated and hence the zigzag detection and JCF would perform like a decode-and-forward decoder.

\section{Concluding Remarks and Future Work}\label{sec:conclude}

The problem of communication in the presence of time delays that cannot be easily compensated, such as what we encounter in asynchronous physical layer network coding, has been studied. There are three main results in the paper. Theorem 10 establishes that fundamentally there is no loss in the information rates achievable in the presence of timing delays which are integer multiples of symbol duration in comparison to the synchronous case. This is the first result to show that integer-valued asynchronism does not cause any reduction in the achievable rates.  This result is obtained through the use of a novel framework called the interleave-deinterleave (IDT) transform in conjunction with quasi-cyclic codes. Secondly, in Section~\ref{sec:higher}, we have shown that delays from the channel can be exploited to decode an increased set of functions at the relays, thereby obtaining higher rates than in the synchronous case in some scenarios.  Finally, an achievable rate in the presence of non-integer valued delays is given in Theorem 13. The rates achievable are higher than those reported earlier in the literature in \cite{najafi13}.

As applications, the IDT-QC codes have been implemented as coding schemes for two channel models, namely the integer-forcing equalized ISI channel and the asynchronous compute-and-forward relay network. For the former, the proposed IDT-QC framework was able to bridge the gap-to-capacity suffered from the cyclic coding scheme recently proposed. For the latter, it has been shown that the proposed IDT-QC scheme not only provides significantly higher rates than the state of the art but also allows the exploitation of the delay dimension which may lead one to rates beyond those achieved by tightly synchronous compute-and-forward. Moreover, practical implementation of the proposed scheme has also been considered where a joint detection and decoding scheme was considered. Simulation results for the two transmitters and one receiver case have further confirmed the theoretical analysis and observations.

Interesting future work includes the following. Theoretically, it is of interest to see how one can further benefit from the delay dimension introduced by the channel. For example, for the symbol-level asynchronous compute-and-forward, it could be the case that some relays are unable to compute any function with a rate above the threshold but some relays are able to compute more than one functions (possibly have the same coefficients but different delays) with rates above the threshold. If the second phase bandwidth is not an issue, the central destination can pick $S$ functions with the highest computation rates, regardless of where the functions are computed. This may lead to higher computation rates than that provided by tightly synchronous compute-and-forward. Practically, it is interesting to design spatially-coupled QC LDPC codes which can further bridge the gap to those theoretical results. Moreover, it is of interest to investigate the impact of using a practical pulse shaping function.

\appendices

\section{Proof of Theorem~\ref{thm:comp_rate_frame}}\label{apx:proof_comp_rate}
 We start by reviewing the nested lattice code of Erez and Zamir \cite{erez04} adopted by \cite{nazer2011CF} for achieving the computation rate in the absence of synchronization errors. Let $\Lambda_f$ and $\Lambda_c$ be two $b$-dimensional lattices with the relationship $\Lambda_c\subseteq\Lambda_f$. A nested lattice code is a code using the minimum-energy coset representatives of $\Lambda_f/\Lambda_c$ as codewords. i.e., $\mc{C}\defeq\Lambda_f \cap \mc{V}_{\Lambda_c}$ where $\mc{V}_{\Lambda}$ is the fundamental Voronoi region of a lattice $\Lambda$. The rate of a nested lattice codes is given by
\begin{equation}
    R = \frac{1}{b}\log\left(\frac{\text{Vol}(\mc{V}_{\Lambda_c})}{\text{Vol}(\mc{V}_{\Lambda_f})}\right).
\end{equation}
Moreover, lattices in \cite{erez04} are constructed by Construction A with $(b,k)$ linear codes over $\mbb{F}_p$; hence, one can further rewrite the rate as $R=k/b\log(p)$. We denote such nested lattice codes as $(b,k)$ nested lattice codes.

Now, let $\mc{C}^{(1)},\mc{C}^{(2)},\ldots,\mc{C}^{(L)}$ be identical $(b,k)$ nested lattice code $\Lambda_f \cap \mc{V}_{\Lambda_c}$ that can achieve $\mc{R}(\mathbf{h}_m,\mathbf{a}_m)$ given in \eqref{eqn:comp_rate_frame} in the absence of synchronization errors. i.e., for a $\varepsilon>0$, there exist sufficiently large $b$ and $p$ such that $k/b\log(p)>\mc{R}(\mathbf{h}_m,\mathbf{a}_m)-\varepsilon$ and $P_e^{(b)}<\varepsilon/L$. Let us again concatenate $L$ codes to form a super-code being the collection of $\mathbf{c}=[\mathbf{c}^1,\mathbf{c}^2,\ldots,\mathbf{c}^L]$. We then feed codewords of the super-code into the IDT transform to freeze bits and add CP. Note that every Construction A lattice can be easily put into a systematic form \cite{Murugan07} and hence freezing bits is feasible.

From \eqref{eqn:y_m_tilde}, one can see that the proposed IDT transform would make the received signal a noisy version of
\begin{align}
    &[a_{m1}\mathbf{x}^{L-\tau_{m1}+1},\ldots,a_{m1}\mathbf{x}^{L-\tau_{m1}}]+\notag  \\  &[a_{m2}\mathbf{x}^{L-\tau_{m2}+1},\ldots,a_{m2}\mathbf{x}^{L-\tau_{m2}}].
\end{align}
Each $b$ sub-block is now a perfectly synchronized compute-and-forward problem. Therefore, we can compute $a_{m1}\mathbf{x}^{l}+a_{m2}\mathbf{x}^{l-\tau_{m2}+\tau_{m1}\hspace{-3pt}\mod L}$ for each $l\in\{1,\ldots,L\}$ separately and use the mapping $\varphi$ to obtain linear combinations in $\mbb{F}_p$. The error probability can be union bounded by $P_e^{Lb}<\varepsilon$ and the actual rate of this strategy is given by \eqref{eqn:R_a_CF}. Now, letting $b,p\rightarrow\infty$ results in vanishing $\varepsilon$ and the rate of each nested lattice sub-code would approach $\mc{R}(\mathbf{h}_m,\mathbf{a}_m)$. Moreover, letting $L\rightarrow\infty$ would make the actual rate converge to the design rate $\mc{R}(\mathbf{h}_m,\mathbf{a}_m)$. This completes the proof.

\bibliographystyle{ieeetr}
\bibliography{journal_abbr,bib_QC_LDPC_IF}

\begin{thebibliography}{10}

\bibitem{michael14}
P.-C. Wang, Y.-C. Huang, and K.~R. Narayanan, ``Asynchronous
  compute-and-forward/integer-forcing with quasi-cyclic codes,'' in {\em Proc.
  IEEE Globecom}, Dec. 2014.

\bibitem{zhang06}
S.~Zhang, S.~C. Liew, and P.~P. Lam, ``Hot topic: Physical-layer network
  coding,'' in {\em Proc. ACM MobiCom}, pp.~358--365, Sept. 2006.

\bibitem{nazer_tutorial}
B.~Nazer and M.~Gastpar, ``Reliable physical-layer network coding,'' {\em
  Proceedings of the IEEE}, vol.~99, pp.~438--460, Mar. 2011.

\bibitem{nazer2011CF}
B.~Nazer and M.~Gastpar, ``Compute-and-forward: Harnessing interference through
  structured codes,'' {\em IEEE Trans. Inf. Theory}, vol.~57, pp.~6463--6486,
  Oct. 2011.

\bibitem{ordentlich12}
O.~Ordentlich and U.~Erez, ``Cyclic-coded integer-forcing equalization,'' {\em
  IEEE Trans. Inf. Theory}, vol.~58, pp.~5804--5815, Sept. 2012.

\bibitem{FuShenli09}
D.~Wang, S.~Fu, and K.~Lu, ``Channel coding design to support asynchronous
  physical layer network coding,'' in {\em Proc. IEEE Globecom}, pp.~1--6, Nov.
  2009.

\bibitem{Lu12}
L.~Lu and S.-C. Liew, ``Asynchronous physical-layer network coding,'' {\em IEEE
  Trans. Wireless Commun.}, vol.~11, pp.~819--831, Feb. 2012.

\bibitem{wu13}
X.~Wu, C.~Zhao, and X.~You, ``Joint {LDPC} and physical-layer network coding
  for asynchronous bi-directional relaying,'' {\em IEEE J. Sel. Areas Commun.},
  vol.~31, pp.~1446--1454, Aug. 2013.

\bibitem{najafi13}
H.~Najafi, M.~O. Damen, and A.~H{\o}rungnes, ``Asynchornous
  compute-and-forward,'' {\em IEEE Trans. Commun.}, vol.~61, pp.~2704--2712,
  July 2013.

\bibitem{qcenc1}
Z.~Li, L.~Chen, L.~Zeng, S.~Lin, and W.~Fong, ``Efficient encoding of
  quasi-cyclic low-density parity-check codes,'' {\em IEEE Trans. Commun.},
  vol.~54, no.~1, pp.~71--81, 2006.

\bibitem{qcdec1}
Z.~Wang and Z.~Cui, ``Low-complexity high-speed decoder design for quasi-cyclic
  {LDPC} codes,'' {\em IEEE Trans. on Very Large Scale Integration Systems},
  vol.~15, no.~1, pp.~104--114, 2007.

\bibitem{qcdec2}
Y.~Chen and K.~Parhi, ``Overlapped message passing for quasi-cyclic low-density
  parity check codes,'' {\em IEEE Trans. on Circuits and Systems I}, vol.~51,
  no.~6, pp.~1106--1113, 2004.

\bibitem{80211n}
``{IEEE} standard for information technology-- local and metropolitan area
  networks-- specific requirements-- part 11: Wireless lan medium access
  control (mac) and physical layer (phy) specifications amendment 5:
  Enhancements for higher throughput,'' {\em {IEEE} Std 802.11n-2009},
  pp.~1--565, 2009.

\bibitem{80216e}
``{IEEE} standard for local and metropolitans area network. part-16; air
  interface for fixes broadband wireless acccess systems,'' {\em {IEEE}
  802.16-2004}, pp.~1--857, 2004.

\bibitem{DVBS2}
``Digital video broadcasting ({DVB}) user guidelines for the second generation
  system for broadcasting, interactive services, news gathering and other
  broadband satellite applications ({DVB-S2}),'' {\em ETSI TR 102 376}, 2005.

\bibitem{YangLiew13}
Q.~Yang and S.~C. Liew, ``Asynchronous convolutional-coded physical-layer
  network coding,'' {\em arXiv:1312.1447v1 [cs.IT]}, Dec. 2013.

\bibitem{Lan07}
L.~Lan, L.~Zeng, Y.~Tai, L.~Chen, S.~Lin, and K.~Abdel-Ghaffar, ``Construction
  of quasi-cyclic ldpc codes for {AWGN} and binary erasure channels: A finite
  field approach,'' {\em IEEE Trans. Inf. Theory}, vol.~53, no.~7,
  pp.~2429--2458, 2007.

\bibitem{CCSD}
{Consultative Committee for Space Data Systems}, ``Low density parity check
  codes for use in near-earth and deep space applications (131.1-o-2 orange
  book),'' Sept. 2007.

\bibitem{Thorpe}
J.~Thorpe, ``Low-density parity-check ({LDPC}) codes constructed from
  protographs,'' in {\em JPL, IPN Progress Report}, pp.~42--154, Aug. 2003.

\bibitem{Feng10}
C.~Feng, D.~Silva, and F.~R. Kschischang, ``An algebraic approach to
  physical-layer network coding,'' {\em IEEE Trans. Inf. Theory}, vol.~59,
  pp.~7576--7596, Nov. 2013.

\bibitem{Engin12}
N.~E. Tunali, K.~Narayanan, J.~Boutros, and Y.-C. Huang, ``Lattices over
  eisenstein integers for compute-and-forward,'' in {\em Proc. Allerton Conf.},
  Oct. 2012.

\bibitem{product_const14}
Y.-C. Huang, K.~Narayanan, and N.~E. Tunali, ``Multistage compute-and-forward
  with multilevel lattice codes based on product constructions,'' {\em
  arXiv:1401.2228 [cs.IT]}, Jan. 2014.

\bibitem{LeechSloane71}
J.~Leech and N.~J.~A. Sloane, ``Sphere packing and error-correcting codes,''
  {\em Canad. J. Math.}, vol.~23, no.~4, pp.~718--745, 1971.

\bibitem{conway1999sphere}
J.~Conway and N.~Sloane, {\em Sphere Packings, Lattices, and Groups}.
\newblock Springer Verlag, 1999.

\bibitem{forney2000}
G.~D. Forney, M.~D. Trott, and S.-Y. Chung, ``Sphere-bound-achieving coset
  codes and multilevel coset codes,'' {\em IEEE Trans. Inf. Theory}, vol.~46,
  pp.~820--850, May 2000.

\bibitem{Guess00}
T.~Guess and M.~Varanasi, ``A new successively decodable coding technique for
  intersymbol-interference channels,'' in {\em Proc. IEEE ISIT}, p.~102, June
  2000.

\bibitem{Gallager}
R.~G. Gallager, {\em Information Theory and Reliable Communication}.
\newblock John Wiley and Sons Inc. New York, NY, USA, 1968.

\bibitem{erez04}
U.~Erez and R.~Zamir, ``Achieving $\tfrac{1}{2}\log(1+\text{{SNR}})$ on the
  {AWGN} channel with lattice encoding and decoding,'' {\em IEEE Trans. Inf.
  Theory}, vol.~50, pp.~2293--2314, Oct. 2004.

\bibitem{Abbasfar07}
A.~Abbasfar, D.~Divsalar, and K.~Yao, ``Accumulate-repeat-accumulate codes,''
  {\em IEEE Trans. Commun.}, vol.~55, pp.~692--702, Apr. 2007.

\bibitem{Huang12}
Q.~Huang, Q.~Diao, S.~Lin, and K.~Abdel-Ghaffar, ``Cyclic and quasi-cyclic
  {LDPC} codes on constrained parity-check matrices and their trapping sets,''
  {\em IEEE Trans. Inf. Theory}, vol.~58, pp.~2648--2671, May 2012.

\bibitem{ModernCodingTheory}
T.~Richardson and R.~Urbanke, {\em Modern Coding Theory}.
\newblock Cambridge University Press, 2008.

\bibitem{pfister01}
H.~Pfister, J.~Soriaga, and P.~Siegel, ``On the achievable information rates of
  finite state {ISI} channels,'' in {\em Proc. IEEE Globecom}, Nov. 2001.

\bibitem{Zhang09}
S.~Zhang and S.-C. Liew, ``Channel coding and decoding in a relay system
  operated with physical-layer network coding,'' {\em IEEE J. Sel. Areas
  Commun.}, vol.~27, pp.~788--796, Feb. 2009.

\bibitem{Murugan07}
K.~Murugan, A;~Azarian and H.~El-Gamal, ``Cooperative lattice coding and
  decoding in half-duplex channels,'' {\em IEEE J. Sel. Areas Commun.},
  vol.~25, pp.~268--279, Feb. 2007.

\end{thebibliography}

\end{document}